\documentclass[twocolumn,showpacs,preprintnumbers,amsmath,amssymb,pre]{revtex4}

\usepackage{graphicx}
\usepackage{dcolumn}
\usepackage{bm}

\begin{document}

\title{Reduction of velocity fluctuations in a turbulent flow of gallium by an external magnetic field.}
\author{Michael Berhanu }
\email{mberhanu@lps.ens.fr}
\author{Basile Gallet}
\author{Nicolas Mordant}
\author{St\'ephan Fauve}

\affiliation{%
Laboratoire de Physique Statistique, Ecole Normale Sup\'erieure \& CNRS, 24 Rue Lhomond, 75231 PARIS Cedex 05, France}%

\date{\today}
\begin{abstract}
The magnetic field of planets or stars is generated by the motion of a conducting fluid through a dynamo instability. The saturation of the magnetic field  occurs through the reaction of the Lorentz force on the flow. In relation to this phenomenon, we
study the effect of a magnetic field on a turbulent flow of liquid
Gallium. The measurement of electric potential differences provides a signal related to the local velocity fluctuations. We
observe a reduction of velocity fluctuations at all frequencies in the spectrum when the magnetic field is increased.
\end{abstract}


\maketitle

Magnetic fields of planets and stars are created by the dynamo instability, which converts part of the kinetic energy of  an electrically conducting fluid  into electromagnetic energy. Recently, several groups succeeded in generating fluid dynamos using liquid sodium flows \cite{Monchaux,Karlsruhe,Gailitis}.  The feedback of the magnetic field on the flow through the Lorentz force  causes the saturation of this instability. This effect is quite well understood in the Riga~\cite{Gailitis} and Karlsruhe~\cite{Karlsruhe} dynamos in terms of mean field magnetohydrodynamics or of the change of the velocity profiles~\cite{Kenjeres,Gailitis,Radler, Tilgner}. In these two cases the turbulence intensity is low and the turbulent motions are restricted to small scales. In the case of the VKS dynamo~\cite{Monchaux} for example, the turbulent fluctuations are present at all scales of the experiment. We are interested in the detailed dynamical mechanism involved in this feedback on the turbulent flow, which remains poorly understood. In particular, it is not clear whether the magnetic field first inhibits the smallest turbulent eddies, i.e. the ones containing the lowest amount of kinetic energy, as first assumed by Batchelor~\cite{Batchelor}, or if it affects all scales of the flow. Several experiments studied the influence of strong magnetic fields on decaying turbulent flows of liquid metal \cite{Sommeria,Alemany,Knaepen,Eckert}. 
Most of them concern freely decaying turbulence and are performed in a range of dimensionless parameters very far from the one involved for weakly supercritical dynamos. Here we focus on the braking effect of the magnetic field on a stationary turbulent flow. A constant magnetic field of tunable amplitude is applied to a gallium flow. The statistics of the velocity field can be accessed through the measurement of electrical potential differences in the metal flow. We report the decrease in the intensity of the velocity fluctuations at all frequencies as the magnetic field is increased.


\begin{figure}
\includegraphics[scale=.3]{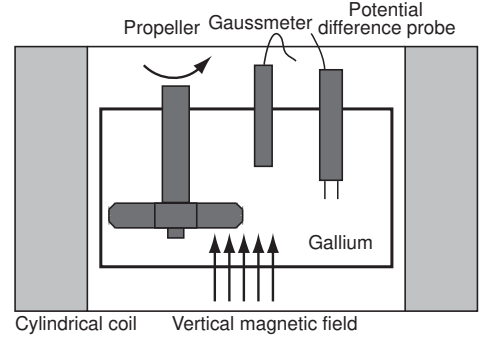}
\caption{\label{coupe3}Cross-section of the experimental setup. A propeller is placed off axis in a cylindrical cell filled with liquid gallium. The vertical magnetic field is generated by a cylindrical coil and is monitored with a gaussmeter. An electrical potential difference is measured by a pair of copper electrodes.}
\end{figure}
One liter of gallium fills a glass cylindrical cell, $12.5$~cm in diameter and $7$~cm in height (fig \ref{coupe3}). A coil imposes a constant vertical magnetic field that can be varied up to 1~kG. 
A propeller is positioned 3 cm off-axis and 3 cm above the bottom of the vessel. Its radius is $R=3.5$~cm. The propeller is rotated so that it ejects the fluid towards the bottom of the cell. The off centering of the propeller prevents  the fluid from rotating as a solid body. This generates a highly fluctuating flow. Rotation frequencies $f_{rot}$ of the propeller range from 2 to 7 Hz.

In the approximations of magnetohydrodynamics, the magnetic field $\mathbf B$ follows the induction equation:
\begin{equation}
\frac{\partial \mathbf B}{\partial t}=\nabla \times (\mathbf v\times \mathbf B)+\eta\Delta \mathbf B\,,
\end{equation}
where $\eta=(\mu_0\sigma)^{-1}$ is the magnetic diffusivity ($\sigma$ is the electrical conductivity) and $\mathbf v$ is the flow velocity.
For an incompressible fluid, the latter is governed by the Navier-Stokes equation with the Lorentz force:
\begin{equation}
\rho\left(\frac{\partial \mathbf v}{\partial t}+(\mathbf v \cdot \nabla)\mathbf v\right)=-\nabla p+\rho\nu \Delta \mathbf v + \mathbf j \times \mathbf B\, ,
\end{equation}
where $\nu$ is the kinematic viscosity, $p$ is the pressure, $\mathbf j$ is the electrical current density and $\rho$ is the density of the fluid.
Three dimensionless numbers can be defined:
(i) the usual hydrodynamic Reynolds number:  $Re=LV/\nu$ ($V$ and $L$ being velocity and length scales), 
(ii) the magnetic Reynolds number $R_m=LV/\eta$  which is the ratio of the induction term over the magnetic diffusive term (Ohmic diffusion) in the induction equation, 
(iii) the interaction parameter $N$ which compares the Lorentz force $\mathbf j \times \mathbf B$ to the convective acceleration in the Navier-Stokes equation. For an applied magnetic field $\mathbf {B_0}$, the induced current is $ \mathbf {j} \sim \sigma (\mathbf {v} \times \mathbf {B_0}) $, so that $N= 
{\sigma L {B_0} ^2}/ {\rho V} $

In the case of gallium one has $\sigma=3.86 \, 10^6$~$\Omega^{- 1}$m$^{- 1}$, $\nu=3.11 \, 10^{- 7}$~m$^2$s$^{- 1}$ and $\rho=6090$~kg/m$^3$. 

In our experiment, the Reynolds number can be varied in the range $5\,10^4<Re<1.7\,10^5$ so that the flow is turbulent. The magnetic Reynolds number is in the range $0.07<R_m<0.25$. The interaction parameter can be rewritten $N=  {\sigma {B_0}^2}/( {2 \pi f_{rot} \rho} )$ by using $R\,f_{rot}$ as a velocity scale.



Velocimetry in liquid metals can be achieved by measuring the electromotive force that arises from the motion of the metal in an applied magnetic field. Measurements of the voltage between two electrodes, a few millimeters apart, have been performed for more than 50 years. The magnetic field can either be a large scale field~\cite{Kolin,Tsinober,Eckert} or the localized field of a small magnet~\cite{Cramer,Ricou}. We use the first configuration in our experiment. The electrodes are located at the end of a cylindrical rod plunging into the liquid metal (fig.~\ref{coupe3}). Each one is made of a 5 mm long copper wire, insulated except at the very end which is in electrical contact with the fluid.  They are separated by a $3.5$~mm wide gap. 

In the approximations of magnetohydrodynamics, Ohm's law in a moving fluid is written : $\,\mathbf{j}=\sigma ( \mathbf{E} + \mathbf{v} \times \mathbf{B}) $. In quasi-static regime one can consider that $\nabla\cdot\mathbf j =0$. With the choice $\nabla\cdot \mathbf A=0$, the divergence of Ohm's law yields~:
\begin{equation}
\Delta \phi=\omega \cdot \mathbf B - \mu_0 \mathbf v \cdot\mathbf j
\label{deltapot}
\end{equation}
The potentials $ \phi$ and $\mathbf A$ are defined by $\mathbf E = -\nabla \phi - \frac{\partial \mathbf A} {\partial t} $ and $\omega$ is the vorticity of the flow. The order of magnitude of the right-hand side terms of (\ref{deltapot}) are~:
$\omega \cdot\mathbf B_0 \sim{B_0 V}/{l_K}$ and  $\mu_0 \mathbf v \cdot \mathbf j \sim{b V}/{l_{\sigma}}$. $b$ is the order of magnitude of the magnetic field induced from $\mathbf B_0$ by the fluid motion. $l_K \sim L {Re}^{-3/4}$ is the Kolmogorov length (dissipative length scale of the velocity) and $l_\sigma \sim L{Rm}^{-3/4}$ is the length scale of ohmic dissipation of the induced currents.  $L$ is chosen as the integral length scale of the flow. In our experiment,  $R_m$ is low so $b\ll B_0$ and the second r.h.s term is negligible so that
\begin{equation}
\Delta \phi= \mathbf \omega \cdot \mathbf B_0
\label{deltapot2}
\end{equation}
Thus potential measurements provide a measurement of the component of vorticity parallel to the  applied field. Similarly to the case of pressure, a direct measurement of $ \phi$ would be nonlocal, because of the Laplacian operator. For length scales larger than the separation $l$ of the electrodes, the measurement of  potential differences gives the gradient of $\phi$.  A signal of order $v B_0$ is obtained (one spatial integration of (\ref{deltapot2})). Thus, we expect the potential difference measurements to be related to the dynamics of the velocity fluctuations, but a precise calibration of the probe is not straightforward. For the fluctuations corresponding to length scales much smaller than $l$, the difference between two decorrelated values of the potential $ \phi$ is measured. We expect the statistics of the potential difference to be the same as that of the potential itself. In the framework of Kolmogorov's theory of turbulence, the Fourier spectrum of vorticity is supposed to behave as $k^{1/3}$. Because of the sweeping of velocity structures by the turbulent flow or the Taylor hypothesis (when relevant \cite{Tennekes}), one gets directly a prediction for the frequency spectrum. We expect the spectrum of the potential difference to behave as $f^{-5/3}$ for frequencies lower than a cutoff frequency $f_c \sim{V}/{2 \pi l} $ and as $f^{-11/3}$ for higher frequencies.


\begin{figure}[!htb]
\centering
\includegraphics[width=8cm]{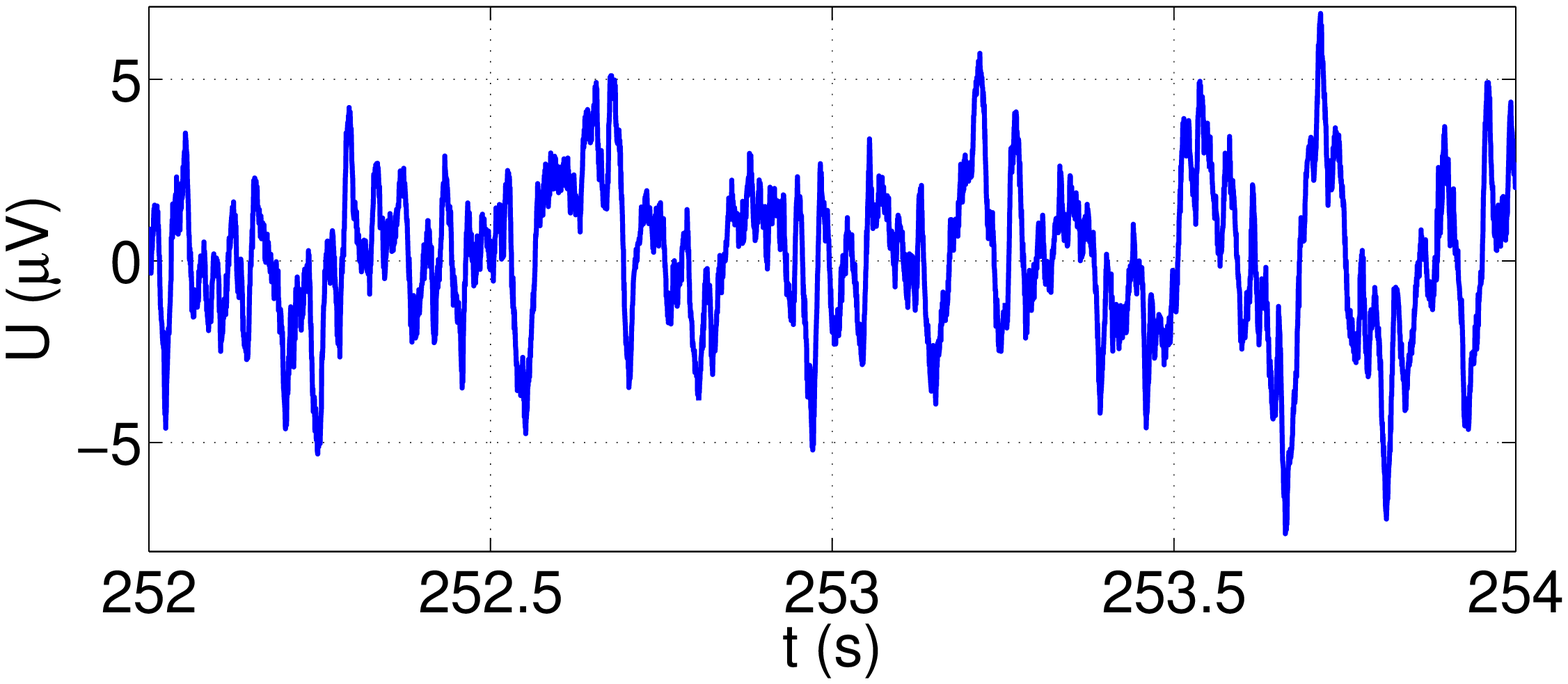}
\includegraphics[width=8cm]{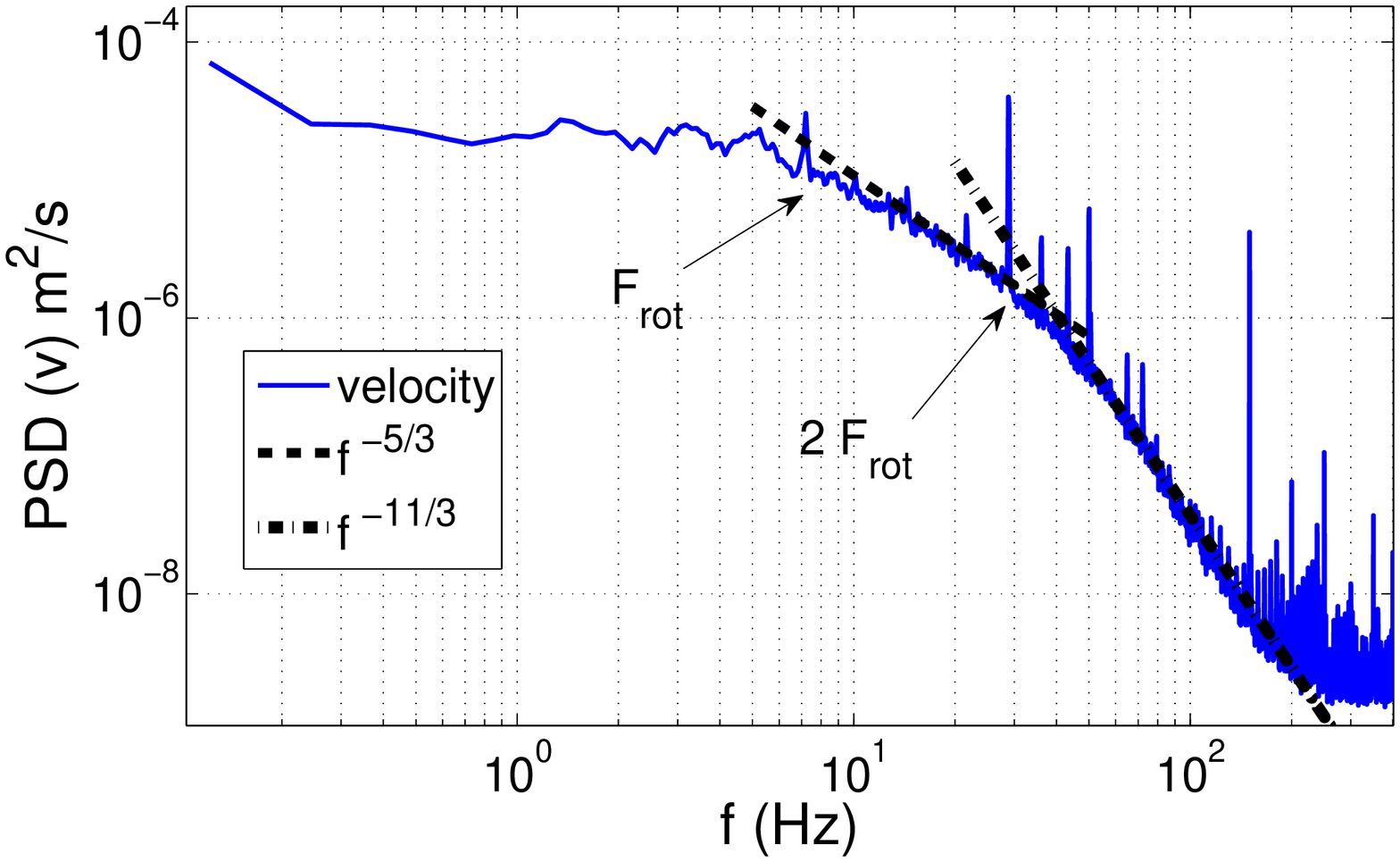}
\caption{Top: temporal signal measured with the potential probe for a rotation frequency of propeller of 7 Hz under a magnetic field of 738 G. Bottom: corresponding power spectrum.}
\label{satfig0}
\end{figure}
For a magnetic field of about 50 Gauss, a signal of a few microvolts is measured. It increases with both the applied field and the rotation frequency of the propeller.
A time series of the potential is displayed in fig.~\ref{satfig0} together with the corresponding temporal power spectrum. The qualitative behavior of the spectrum is in agreement with the discussion of the previous paragraph. For intermediate values of the frequencies, starting from the rotation frequency of the propeller, the spectrum decays following roughly a power law which exponent is close to $-5/3$. For frequencies larger than 40 Hz, a steeper power law decay is observed, with an exponent close to $-11/3$. Beyond 100 Hz, the signal falls under the noise level. This is consistent with a hydrodynamical interpretation of our potential measurements. For lower driving speeds, the $f^ {- 5/3} $ zone is not very visible, most likely because of the lower value of the Reynolds number that may not develop a clear inertial range.  The quantitative behavior of the spectra depends somewhat on the position and the orientation of the probe, because of the inhomogeneity of the flow.

\begin{figure}
\includegraphics[width=8cm]{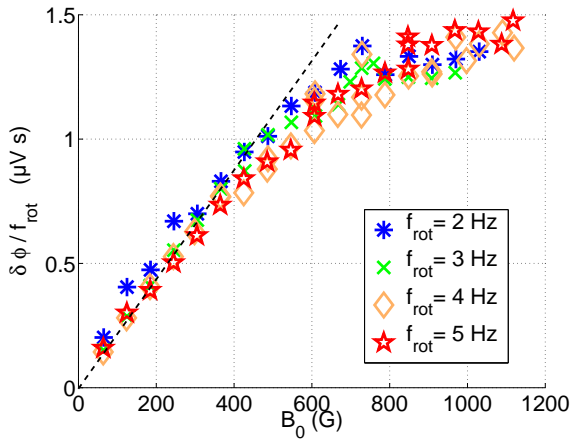}
\caption{\label{satfig3} Standard deviation of the fluctuations of the potential difference divided by the rotation frequency of the propeller versus applied magnetic field .}
\end{figure}
We evaluate quantitatively the amplitude of the fluctuations of the potential difference $\delta\phi$ by the standard deviation $\sigma_{\delta\phi}$ of the recorded signal. As seen in fig.~\ref{satfig3},
when the applied magnetic field  is gradually increased, the amplitude of $\delta\phi$ first grows linearly with $B_0$ and $f_{rot}$. For larger values of the magnetic field, a departure from the linear behavior in $B_0$ is observed which we ascribe to the effect of the Lorentz force on the flow.

\begin{figure}
\includegraphics[width=8.5cm]{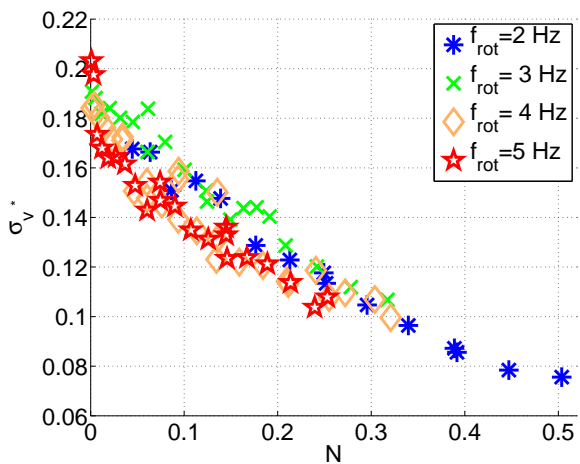}
\caption{\label{figtot} Standard deviation of dimensionless velocity fluctuations, versus interaction parameter N.}
\end{figure}
The potential difference is divided by $B_0$ (constant during a given measurement) and also by the width $l$ of the probe.  This quantity $v={\delta\phi}/B_0l$ is homogeneous to a velocity and is related to the velocity fluctuations in the vicinity of the probe. Although the link is not direct between velocity and potential difference, for simplicity, we call $v$ (somewhat abusively) ``velocity" in the following. The standard deviation of this velocity is non dimensionalized by $Rf_{rot}$ and is noted $\sigma_{v^*}=\sigma_v/Rf_{rot} $ in the following. Within the accuracy of our measurements, it decreases linearly with the interaction parameter, as seen in fig.~\ref{figtot}. The decay of $ \sigma_{v^*} $ can reach a factor 2.

The obtained curves look similar to the potential measurements at large scale performed by Steenbeck et al. \cite{Steenbeck}. They recorded the potential produced by $\alpha$ effect at the border of  a dedicated device with sodium flow. In their experiment, the average potential decreases with the applied field, and follows roughly a $1/N$ decay law. Their measurements are interpreted as a quenching of the $ \alpha$ effect, corresponding to a braking of the flow. This saturates the efficiency of  the induction processes and is responsible for the saturation of the $\alpha$-dynamo \cite{Dormy}. It supports the model of the $ \alpha$ effect, where the average electromotive force is written  as $ \langle\mathbf{v} \times \mathbf{B} \rangle = \alpha \mathbf B_0 $. Here we observe directly the reduction of the turbulent velocity fluctuations resulting from the magnetic braking of the flow.

\begin{figure}
\includegraphics[width=8.5cm]{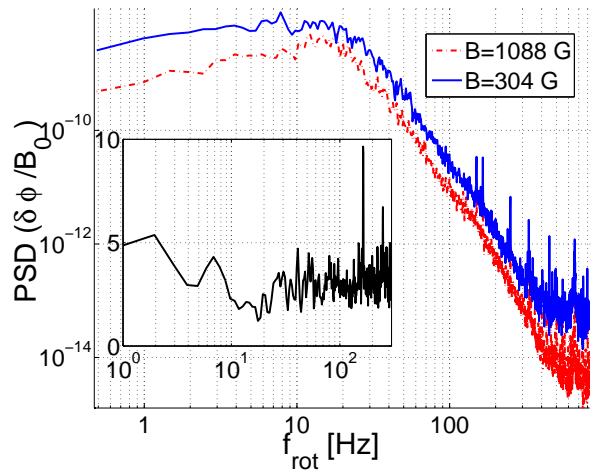}
\caption{\label{spectre5sat} Comparison  between power spectra of velocity  for a rotation frequency of 5 Hz for two different values of the external magnetic field. Inset : ratio between the two power spectrum densities.}
\end{figure}

One open question on the spectral behavior of the magnetic braking phenomenon is whether it acts on a privileged scale. In particular, one could expect the braking to be more efficient at the smallest scales of the flow and therefore to damp preferentially the highest frequencies of the spectrum. Indeed, following Batchelor~\cite{Batchelor}, a scale $\delta$ would be damped if the volumic kinetic energy $\frac{1}{2} \rho v^2_\delta$ at this scale is below the volumic magnetic energy $\frac{1}{2} {{B_0}^2}/{\mu_0}$. The braking effect of the magnetic field could then be modeled as an enhanced dissipation. To answer this question, we compare the spectra of $\delta\phi/B_0$ for two values of the applied magnetic field at a given frequency $f_{rot}$ of the propeller (fig.~\ref{spectre5sat}). The spectrum corresponding to the highest $B_0$ is below the other one but no relative variation in the damping is observed over the frequencies. The two spectra can be obtained one from another by a simple scaling factor which is the ratio of the two squared standard deviations $\sigma_{v^*}^2$. The magnetic braking is performed similarly at all scales in the present range of dimensionless parameters.


Several authors \cite{Alemany,Sommeria,Knaepen,Eckert} studied the influence of a magnetic field on a flow both experimentally and numerically. For decaying turbulence, they observed a faster damping of the velocity field for large values of the interaction parameter $N$. This occurs simultaneously with a bidimensionnalisation of the flow. In our experiment, no evolution is observed toward a $f^{- 3} $ decay of the spectrum, expected for 2D turbulence. One difference may be due to the continuous forcing in our experiment but also to the different values of $N$ and $Rm$.  In this case the large scale eddy turnover time may be lower than the decay time of the velocity field via the Lorentz force, for moderate values of $N$. 
The mechanical energy injection $ \epsilon_0 \sim{\rho V^3}/{L} $ should be compared to the power of the Lorentz force. The latter is identified with the dissipation by Joule effect $\mathbf j\cdot \mathbf E \sim \sigma V^2 {B_0}^2$. The ratio of the two is $N$ which takes values lower than one in our case. The dissipation of the kinetic energy is thus done primarily via the turbulent cascade at $N<1$. The current density is of order $ \mathbf j \sim \sigma \mathbf V\times \mathbf B_0 $, so that the Lorentz force is $\sigma {B_0}^2 V $.
Thus we  can build a  braking time $ \tau_m= \rho/{\sigma {B_0}^2} $, which must be compared to the eddy turnover time $ \tau_h \sim \frac{L}{v} $ - after which the velocity field is decorrelated. Once again, for $N$ smaller than unity, one has $\tau_h/\tau_m=N<1$, which means that the magnetic field does not have enough time to affect the structure of the flow and to make it anisotropic.
Consequently in the range of parameters of our experiment and also for the saturation of the dynamo effect, the application of a magnetic field to a turbulent flow, brings an additional braking term, without changing the spectral behavior of velocity fluctuations. On the contrary, for $N>1$, a variation of the slope of the velocity spectrum with the applied field is expected -- as observed by Eckert {\it et al.}~\cite{Eckert}.
The total dissipation of the flow increases with the magnetic field because of Joule dissipation. A smaller part of the injected power is available for the turbulent cascade. Nevertheless it neither changes much of its dynamics nor breaks its scale invariance. 
An efficiency factor $F<1$ depending on the interaction parameter can be introduced to estimate the effective kinetic dissipation rate. From the usual arguments of Kolmogorov's 1941 theory, the velocity spectrum can then be written $E(k)\sim \left( {V^3}{L}^{-1} F(N)  \right)^{2/3} k^{-5/3} $. Consequently when the magnetic field is increased, the velocity spectrum preserves its shape, but its amplitude decreases as $F(N)^{2/3}$.

In summary, we observed the influence of the Lorentz force on a turbulent flow of gallium through measurements of electrical potential differences. A strong reduction of the turbulent fluctuations is evidenced as the applied magnetic field is increased. In the range of interaction parameter and magnetic Reynolds number corresponding to our measurements, the observed spectra show that the dynamics of turbulence is affected by the presence of the magnetic field only through a scaling efficiency factor. 

\begin{acknowledgments}
We acknowledge  stimulating discussions on the work reported in this article with S\'ebastien Auma\^itre, Fran\c cois P\'etr\'elis and Claudio Falc\'on.
\end{acknowledgments}

\bibliography{redufluc_resub}

\begin{thebibliography}{18}
\expandafter\ifx\csname natexlab\endcsname\relax\def\natexlab#1{#1}\fi
\expandafter\ifx\csname bibnamefont\endcsname\relax
  \def\bibnamefont#1{#1}\fi
\expandafter\ifx\csname bibfnamefont\endcsname\relax
  \def\bibfnamefont#1{#1}\fi
\expandafter\ifx\csname citenamefont\endcsname\relax
  \def\citenamefont#1{#1}\fi
\expandafter\ifx\csname url\endcsname\relax
  \def\url#1{\texttt{#1}}\fi
\expandafter\ifx\csname urlprefix\endcsname\relax\def\urlprefix{URL }\fi
\providecommand{\bibinfo}[2]{#2}
\providecommand{\eprint}[2][]{\url{#2}}

\bibitem[{\citenamefont{Monchaux et~al.}(2007)\citenamefont{Monchaux, Berhanu,
  Bourgoin, Moulin, Odier, Pinton, Volk, Fauve, Mordant, Petrelis
  et~al.}}]{Monchaux}
\bibinfo{author}{\bibfnamefont{R.}~\bibnamefont{Monchaux}},
  \bibinfo{author}{\bibfnamefont{M.}~\bibnamefont{Berhanu}},
  \bibinfo{author}{\bibfnamefont{M.}~\bibnamefont{Bourgoin}},
  \bibinfo{author}{\bibfnamefont{M.}~\bibnamefont{Moulin}},
  \bibinfo{author}{\bibfnamefont{P.}~\bibnamefont{Odier}},
  \bibinfo{author}{\bibfnamefont{J.-F.} \bibnamefont{Pinton}},
  \bibinfo{author}{\bibfnamefont{R.}~\bibnamefont{Volk}},
  \bibinfo{author}{\bibfnamefont{S.}~\bibnamefont{Fauve}},
  \bibinfo{author}{\bibfnamefont{N.}~\bibnamefont{Mordant}},
  \bibinfo{author}{\bibfnamefont{F.}~\bibnamefont{Petrelis}},
  \bibnamefont{et~al.}, \bibinfo{journal}{Phys. Rev. Lett.}
  \textbf{\bibinfo{volume}{98}}, \bibinfo{pages}{044502}
  (\bibinfo{year}{2007}).

\bibitem[{\citenamefont{Stieglitz and M\"uller}(2001)}]{Karlsruhe}
\bibinfo{author}{\bibfnamefont{R.}~\bibnamefont{Stieglitz}} \bibnamefont{and}
  \bibinfo{author}{\bibfnamefont{U.}~\bibnamefont{M\"uller}},
  \bibinfo{journal}{Phys. Fluids} \textbf{\bibinfo{volume}{13}},
  \bibinfo{pages}{561} (\bibinfo{year}{2001}).

\bibitem[{\citenamefont{Gailitis et~al.}(2004)\citenamefont{Gailitis,
  Lielausis, Platacis, Gerbeth, and Stefani}}]{Gailitis}
\bibinfo{author}{\bibfnamefont{A.}~\bibnamefont{Gailitis}},
  \bibinfo{author}{\bibfnamefont{O.}~\bibnamefont{Lielausis}},
  \bibinfo{author}{\bibfnamefont{E.}~\bibnamefont{Platacis}},
  \bibinfo{author}{\bibfnamefont{G.}~\bibnamefont{Gerbeth}}, \bibnamefont{and}
  \bibinfo{author}{\bibfnamefont{F.}~\bibnamefont{Stefani}},
  \bibinfo{journal}{Physics of Plasmas} \textbf{\bibinfo{volume}{11}},
  \bibinfo{pages}{2838} (\bibinfo{year}{2004}).

\bibitem[{\citenamefont{Kenjere\v{s} et~al.}(2006)\citenamefont{Kenjere\v{s},
  Hanjali\'c, Stefani, Gerbeth, and Gailitis}}]{Kenjeres}
\bibinfo{author}{\bibfnamefont{S.}~\bibnamefont{Kenjere\v{s}}},
  \bibinfo{author}{\bibfnamefont{K.}~\bibnamefont{Hanjali\'c}},
  \bibinfo{author}{\bibfnamefont{F.}~\bibnamefont{Stefani}},
  \bibinfo{author}{\bibfnamefont{G.}~\bibnamefont{Gerbeth}}, \bibnamefont{and}
  \bibinfo{author}{\bibfnamefont{A.}~\bibnamefont{Gailitis}},
  \bibinfo{journal}{Physics of Plasmas} \textbf{\bibinfo{volume}{13}},
  \bibinfo{pages}{122308} (\bibinfo{year}{2006}).

\bibitem[{\citenamefont{R\"adler et~al.}(2002)\citenamefont{R\"adler,
  Rheinhardt, Apstein, and Fuchs}}]{Radler}
\bibinfo{author}{\bibfnamefont{K.-H.} \bibnamefont{R\"adler}},
  \bibinfo{author}{\bibfnamefont{M.}~\bibnamefont{Rheinhardt}},
  \bibinfo{author}{\bibfnamefont{E.}~\bibnamefont{Apstein}}, \bibnamefont{and}
  \bibinfo{author}{\bibfnamefont{H.}~\bibnamefont{Fuchs}},
  \bibinfo{journal}{Magnetohydrodynamics} \textbf{\bibinfo{volume}{38}},
  \bibinfo{pages}{73} (\bibinfo{year}{2002}).

\bibitem[{\citenamefont{Tilgner and Busse}(2002)}]{Tilgner}
\bibinfo{author}{\bibfnamefont{A.}~\bibnamefont{Tilgner}} \bibnamefont{and}
  \bibinfo{author}{\bibfnamefont{F.~H.} \bibnamefont{Busse}},
  \bibinfo{journal}{Magnetohydrodynamics} \textbf{\bibinfo{volume}{38}},
  \bibinfo{pages}{35} (\bibinfo{year}{2002}).

\bibitem[{\citenamefont{Batchelor}(1950)}]{Batchelor}
\bibinfo{author}{\bibfnamefont{G.~K.} \bibnamefont{Batchelor}},
  \bibinfo{journal}{Proc. Roy. Soc. London A} \textbf{\bibinfo{volume}{201}},
  \bibinfo{pages}{1066} (\bibinfo{year}{1950}).

\bibitem[{\citenamefont{Sommeria and Moreau}(1982)}]{Sommeria}
\bibinfo{author}{\bibfnamefont{J.}~\bibnamefont{Sommeria}} \bibnamefont{and}
  \bibinfo{author}{\bibfnamefont{R.}~\bibnamefont{Moreau}},
  \bibinfo{journal}{J. Fluid Mech.} \textbf{\bibinfo{volume}{118}},
  \bibinfo{pages}{507} (\bibinfo{year}{1982}).

\bibitem[{\citenamefont{Alemany et~al.}(1979)\citenamefont{Alemany, Moreau,
  Sulem, and Frisch}}]{Alemany}
\bibinfo{author}{\bibfnamefont{A.}~\bibnamefont{Alemany}},
  \bibinfo{author}{\bibfnamefont{R.}~\bibnamefont{Moreau}},
  \bibinfo{author}{\bibfnamefont{R.~L.} \bibnamefont{Sulem}}, \bibnamefont{and}
  \bibinfo{author}{\bibfnamefont{U.}~\bibnamefont{Frisch}},
  \bibinfo{journal}{J. de Mecanique} \textbf{\bibinfo{volume}{18}},
  \bibinfo{pages}{277} (\bibinfo{year}{1979}).

\bibitem[{\citenamefont{Knaepen and Moreau}(2008)}]{Knaepen}
\bibinfo{author}{\bibfnamefont{B.}~\bibnamefont{Knaepen}} \bibnamefont{and}
  \bibinfo{author}{\bibfnamefont{R.}~\bibnamefont{Moreau}},
  \bibinfo{journal}{Ann. Rev. Fluid Mech.} \textbf{\bibinfo{volume}{40}},
  \bibinfo{pages}{25} (\bibinfo{year}{2008}).

\bibitem[{\citenamefont{Eckert et~al.}(2001)\citenamefont{Eckert, Gerbeth,
  Witke, and Langenbrunner}}]{Eckert}
\bibinfo{author}{\bibfnamefont{S.}~\bibnamefont{Eckert}},
  \bibinfo{author}{\bibfnamefont{G.}~\bibnamefont{Gerbeth}},
  \bibinfo{author}{\bibfnamefont{W.}~\bibnamefont{Witke}}, \bibnamefont{and}
  \bibinfo{author}{\bibfnamefont{H.}~\bibnamefont{Langenbrunner}},
  \bibinfo{journal}{Int. J. Heat Mass Transfer} \textbf{\bibinfo{volume}{22}},
  \bibinfo{pages}{358} (\bibinfo{year}{2001}).

\bibitem[{\citenamefont{Kolin}(1943)}]{Kolin}
\bibinfo{author}{\bibfnamefont{A.}~\bibnamefont{Kolin}}, \bibinfo{journal}{J.
  Appl. Phys.} \textbf{\bibinfo{volume}{15}}, \bibinfo{pages}{150}
  (\bibinfo{year}{1943}).

\bibitem[{\citenamefont{Tsinober et~al.}(1987)\citenamefont{Tsinober, Kit, and
  Teitel}}]{Tsinober}
\bibinfo{author}{\bibfnamefont{A.}~\bibnamefont{Tsinober}},
  \bibinfo{author}{\bibfnamefont{E.}~\bibnamefont{Kit}}, \bibnamefont{and}
  \bibinfo{author}{\bibfnamefont{M.}~\bibnamefont{Teitel}},
  \bibinfo{journal}{J. Fluid Mech.} \textbf{\bibinfo{volume}{175}},
  \bibinfo{pages}{447} (\bibinfo{year}{1987}).

\bibitem[{\citenamefont{Cramer et~al.}(2006)\citenamefont{Cramer, Varshney,
  Gundrum, and Gerbeth}}]{Cramer}
\bibinfo{author}{\bibfnamefont{A.}~\bibnamefont{Cramer}},
  \bibinfo{author}{\bibfnamefont{K.}~\bibnamefont{Varshney}},
  \bibinfo{author}{\bibfnamefont{T.}~\bibnamefont{Gundrum}}, \bibnamefont{and}
  \bibinfo{author}{\bibfnamefont{G.}~\bibnamefont{Gerbeth}},
  \bibinfo{journal}{Flow Meas. Instr.} \textbf{\bibinfo{volume}{17}},
  \bibinfo{pages}{1} (\bibinfo{year}{2006}).

\bibitem[{\citenamefont{Ricou and Viv\`es}(1982)}]{Ricou}
\bibinfo{author}{\bibfnamefont{R.}~\bibnamefont{Ricou}} \bibnamefont{and}
  \bibinfo{author}{\bibfnamefont{C.}~\bibnamefont{Viv\`es}},
  \bibinfo{journal}{Int. J. Heat Mass Transfer} \textbf{\bibinfo{volume}{25}},
  \bibinfo{pages}{1579} (\bibinfo{year}{1982}).

\bibitem[{\citenamefont{Tennekes and Lumley}(1972)}]{Tennekes}
\bibinfo{author}{\bibfnamefont{H.}~\bibnamefont{Tennekes}} \bibnamefont{and}
  \bibinfo{author}{\bibfnamefont{J.~L.} \bibnamefont{Lumley}},
  \emph{\bibinfo{title}{A first course in Turbulence}} (\bibinfo{publisher}{The
  MIT press}, \bibinfo{year}{1972}).

\bibitem[{\citenamefont{Steenbeck et~al.}(1968)\citenamefont{Steenbeck, Kirko,
  Gailitis, Klyavinya, Kraus, Laumanis, and Lielausis}}]{Steenbeck}
\bibinfo{author}{\bibfnamefont{M.}~\bibnamefont{Steenbeck}},
  \bibinfo{author}{\bibfnamefont{I.~M.} \bibnamefont{Kirko}},
  \bibinfo{author}{\bibfnamefont{A.}~\bibnamefont{Gailitis}},
  \bibinfo{author}{\bibfnamefont{A.~P.} \bibnamefont{Klyavinya}},
  \bibinfo{author}{\bibfnamefont{F.}~\bibnamefont{Kraus}},
  \bibinfo{author}{\bibfnamefont{I.~Y.} \bibnamefont{Laumanis}},
  \bibnamefont{and} \bibinfo{author}{\bibfnamefont{O.~A.}
  \bibnamefont{Lielausis}}, \bibinfo{journal}{Soviet Physics - Doklady}
  \textbf{\bibinfo{volume}{13}}, \bibinfo{pages}{443} (\bibinfo{year}{1968}).

\bibitem[{\citenamefont{Dormy and Soward}(2007)}]{Dormy}
\bibinfo{author}{\bibfnamefont{E.}~\bibnamefont{Dormy}} \bibnamefont{and}
  \bibinfo{author}{\bibfnamefont{A.~M.} \bibnamefont{Soward}},
  \emph{\bibinfo{title}{Mathematical aspects of Natural Dynamos}}
  (\bibinfo{publisher}{Cambridge University Press}, \bibinfo{year}{2007}).

\end{thebibliography}

\end{document}